# Demystifying the use of Compression in Virtual Production

**Dr. Anil Kokaram**

Sigmedia Group, Department of Electronic and Electrical Engineering, Trinity College Dublin, Ireland. anil.kokaram@tcd.ie

**Vibhoothi Vibhoothi**

Sigmedia Group, Department of Electronic and Electrical Engineering, Trinity College Dublin, Ireland. vibhootv@tcd.ie

**Julien Zouein**

Sigmedia Group, Department of Electronic and Electrical Engineering, Trinity College Dublin, Ireland. zoueinj@tcd.ie

**Dr. Francois Pitie**

Sigmedia Group, Department of Electronic and Electrical Engineering, Trinity College Dublin, Ireland. pitief@tcd.ie

**Christopher Nash**

Disguise, London, UK chris.nash@disguise.one

**James Bentley**

Disguise, London, UK james.bentley@disguise.one

**Philip Coulam-Jones**

Disguise, London, UK philip.coulam-jones@disguise.one

**Abstract.** *Virtual Production (VP) technologies have continued to improve the flexibility of on-set filming and enhance the live concert experience. The core technology of VP relies on high-resolution, high-brightness LED panels to playback/render video content. There are a number of technical challenges to effective deployment e.g. image tile synchronisation across the panels, cross panel colour balancing and compensating for colour fluctuations due to changes in camera angles. Given the complexity and potential quality degradation, the industry prefers "pristine" or lossless compressed source material for displays, which requires significant storage and bandwidth. Modern lossy compression standards like AV1 or H.265 could maintain the same quality at significantly lower bitrates and resource demands. There is yet no agreed methodology for assessing the impact of these standards on quality when the VP scene is recorded in-camera. We present a methodology to assess this impact by comparing lossless and lossy compressed footage displayed through VP screens and recorded in-camera. We assess the quality impact of HAP/NotchLC/Daniel2 and AV1/HEVC/H.264 compression bitrates from 2 Mb/s to 2000 Mb/s with various GOP sizes. Several perceptual quality metrics are then used to automatically evaluate in-camera picture quality, referencing the original uncompressed source content through the LED wall. Our results show that we can achieve the same quality with hybrid codecs as with intermediate encoders at orders of magnitude less bitrate and storage requirements.*

# Introduction

Virtual production (VP) is a broad term referring to a spectrum of computer-aided production and visualization filmmaking methods [1] [2] [3]. VP can refer to the infrastructure allowing production teams to collaboratively create in real-time and remotely. It can also refer to the process of simulating environments on-set using appropriate background scene rendering, implicit lighting and camera management. Both developments have contributed to a far richer and potentially more cost effective environment for cinema content production. In this paper we are specifically addressing the VP concept in terms of simulated environments on-set. The sophistication of these sets has grown enormously, built around the foundations set by the work of the ASC Committee on Virtual Production since 2009 [4] and Debevec et al [5]. These workflows are typically highly complex and require sophisticated hardware and software for rendering. The complexity of integrating these technologies has led to the SMPTE On-Set Virtual Production Initiative [6], which aims to standardise and clarify VP practices.

A key aspect contributing to this complexity is that the background plates needed for the LED walls are typically high resolution and high dynamic range. As they are captured directly off of the panels, and recorded in-camera it is accepted lore that the images must be of pristine quality. On the face of it, that means raw pixel formats. But raw formats create too much data to enable real time playback from practical storage and distribution hardware. Using compression is therefore inevitable and dedicated hardware is typically employed to enable real-time playback with minimal latency. The requirements of real time playback and low latency have encouraged the industry to consider near-lossless proprietary compression systems accelerated by GPUs as the only available route forward. However, existing video compression formats HEVC and AV1 provide orders of magnitude lower bitrate with high picture quality. In addition, existing GPU hardware already contains hardware encoders/decoders which enable real time and low latency playback at high resolution with these standards.

In this paper, we compare the use of a variety of compression formats in providing content to LED panels for VP. In so doing we introduce an evaluation setup for comparing picture quality on-set using these codecs and show that x20 reduction in bitrate is possible without loss of quality. We do not conduct subjective tests in this work but we use perceptual metrics to capture similar information.

# Background

## *Virtual Production*

In its current incarnation, the use of VP on-set involves the use of high brightness LED panels to render environment scenery. In a most basic sense this reimagines the use of backprojection for compositing scene elements in camera, pioneered in the 1950's. A live foreground scene is composited with a pre-recorded background scene by recording the entire composition live, in studio. This replaces the use of a green screen for compositing "in post" and instead compositing takes place on set itself. In the making of the movie "Gravity" for instance (2013), the environment and background scene around the actors was created with an LED panel volume, shaped into a cube of side 6m [7]. The key challenge is to align the geometry of the recorded footage and the in-studio setup including the camera, significantly increasing the complexity of pre-production recording. The use of completely synthetic projected imagery simplifies this problem by using a live renderer to take the in-studio geometry and camera viewpoint into account. That more complicated system is enabled by live renderers like Unreal Engine's nDisplay [8] for example. This then allows in-camera visual effects to be generated. One popular visual effect is the "virtual set extension" in which extended imagery outside the boundary of the captured footage can be generated live.

Other than the advantage of reducing travel budgets in shooting, there are significant picture quality advantages of VP with LED walls [9] [3]. Lighting cast by the displays is bright enough to provide more natural lighting environments for the studio capture and the "green cast" due to reflection of a green screen is no longer an issue. In addition many of these elements are now programmable live on set enabling real time visualisation and feedback with the Director of Photography.

In more recent developments, VP has been deployed in live concert environments for background VFX and interactive video that creates immersive environments for audiences and performers alike. In that scenario the live audience perceives the picture quality directly rendered on stage, while remote viewers perceive picture quality rendered by cameras at the venue.

As VP technology evolves [10] [11], there is a need to measure the differences between picture quality rendered in pre-production or by the live game engine renderer, and the picture quality rendered in-camera. Without some objective measure of this difference, systems default to using lossless compression or raw image formats for transport of images to the LED walls. This leads to very large file sizes for storage and transport of media. At bitrates about 2 Gbps the use of these intermediate codecs (HAP, NotchLC, Daniel2) leads to 15 GBytes for a minute of footage, when a hybrid codec could render the same quality at 20 Mbps which is just 150 Mbytes. Intuitively, there are so many elements of the LED renderer and the camera capture which can cause image degradation e.g. aliasing in capture, colour rendering artefacts due to camera viewpoints, that the impact of the quality of the data source might be much less than expected.

It is generally also believed that employing lossy compression formats necessarily implies an increase in latency in rendering images. Latency reduces the ability of a system to deliver real time visualization. Latency in lossy compression standards is a corollary of the ability to deliver high compression rates through inter-frame prediction across multiple frames. The decoder must receive the frames before rendering an image. This is certainly true in general, but there are many modes of operation of an encoder, and it is possible to achieve real time decoding with no latency if the right mode is chosen. We discuss these issues next.

### *Video Compression*

Video compression is an essential component of any modern video streaming pipeline. Over the years, compression techniques have evolved, with modern (hybrid) codecs like AV1 from the Alliance for Open Media (AOM) or HEVC from MPEG offering orders of magnitude bitrate savings over older encoders such as H.264. In VP, where multiple assets, scenes, and overlays are rendered in real time, achieving efficient compression and rendering is challenging [12]. Three main issues are of concern when considering modern video compression tools in VP pipelines: speed, latency and picture quality. In this study we compare a selection of popular codecs as follows.

- **H.264/AVC**: Standardized in 2004, AVC is a widely adopted lossy compression format used in most internet applications, supporting resolutions up to UHD-2 (8K). However, hardware implementations often limit features; for example, NVIDIA GPUs do not support 10-bit colour depth in AVC.
- **H.265/HEVC**: Standardized in 2013, HEVC is a lossy format that natively supports 10-bit profiles, offering greater efficiency than AVC.
- **AV1**: Standardised in 2018, AV1 is a royalty-free successor to VP9, providing roughly 30% more coding efficiency than HEVC. It is integrated into the latest NVIDIA, AMD, and Intel GPUs but is currently limited to YUV 4:2:0 at 10 bits. AV1 is a traditional block-based encoder in many respects but has targeted high quality compression with mandatory film-grain synthesis as part of the standard.
- **HAP**: An open-source codec from Vidvox Ltd., HAP leverages GPU acceleration for fast encoding and decoding. For higher quality (HAP-Q) the codec uses YCoCg-R [13] colour format. It is comparable in file size to ProRes 422.
- **NotchLC**: Developed by 10Bit Ltd., NotchLC is a free high-quality (not open source) GPU powered intermediate codec that uses LZ4 compression to achieve compression. Uniquely, it uses 12bits for Luma and 8 bits for chroma.
- **Daniel2**: Released in 2017 by the Cingey team [14], the Digital Animation Encoder Library 2 (Daniel2) is a GPU-based intermediate codec based on Discrete Cosine Transforms (DCT) for compression coupled with a simple lossless data compression technique [15].

HAP, NotchLC and Daniel2 are all intermediate compression formats (perceptually lossless) popular in the VP community. They are not open standards but guarantee that the quality of the original footage remains untouched in decoding. The other codecs in our comparison set are all lossy compression formats (hybrid codecs) used for video streaming and broadcast television. These other codecs are all open in the sense that the technologies are well understood and described in standards documents. As far as picture quality is concerned, every one of the lossy compression formats can be configured to be perceptually lossless by choosing the right encoding parameters. In our experiments we investigate a range of these parameters (quantiser setting).

On the speed issue, all the formats are decodable in real time. We employ the hardware implementation of these codecs as integrated into the NVIDIA range of GPUs and accessible with the "nvenc" tool in software. **Figure 1** shows the mean encoding time (frames encoded per second) for a variety of codecs tested across our dataset (presented later). As can be seen, the current intermediate encoding formats used in VP (HAP, NotchLC and Daniel2) actually take more time to encode the clips than the hardware invocation of the other compression formats we examine in our test setup.

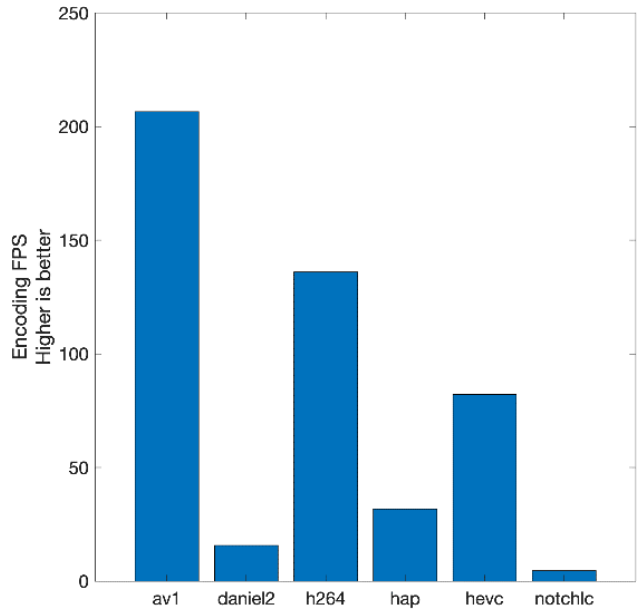

**Figure 1:** Mean Encoding Time (FPS) for different codecs over our test dataset (UHD-1 resolution achieving the same VMAF quality level 100). Using the Nvidia GPU our hybrid codecs (H.264, HEVC and AV1) can achieve more than 50 fps encoding in any mode (intra or non-intra).

The final issue of latency is addressed by investigating different encoding modes for the lossy codecs. Each of these codecs perform compression on the motion compensated frame difference between frames instead of compressing the raw pixel data alone. Therefore each encoder must set the number of frames over which motion compensation is allowed to take place. That group of frames is called a Group of Pictures (GOP). We examine in this study GOP sizes of 0 (implying zero latency in decoding and no inter-frame prediction) up to the entire length of the clip.

### *Video Quality Assessment*

Porrmann [16] evaluated several intermediate codecs (ProRes, HAP, DivX, DPX, NotchLC, Daniel2) for post-production workflows using PSNR and SSIM. Daniel2 performed best, followed by DNxHR-444 and NotchLC. This was outside of the VP display/capture context. To enable an assessment of picture quality in the context of VP [19], and taking into account perceptual quality [17] [18], we use the following metrics in addition to PSNR.

- i) *VMAF*: Video Multimethod Assessment Fusion [20] [21] is a perceptually relevant metric developed by Netflix. That metric computes different objective features of a video to account for both spatial and temporal behaviour. It has been widely adopted in the industry streaming community and research in compression.
- ii) *ColorVideoVDP*: CVVDP [22] is a video and image quality metric that models spatial and temporal aspects of vision for both luminance and colour based on psychophysical observations. This was targeted at XR display artefacts.

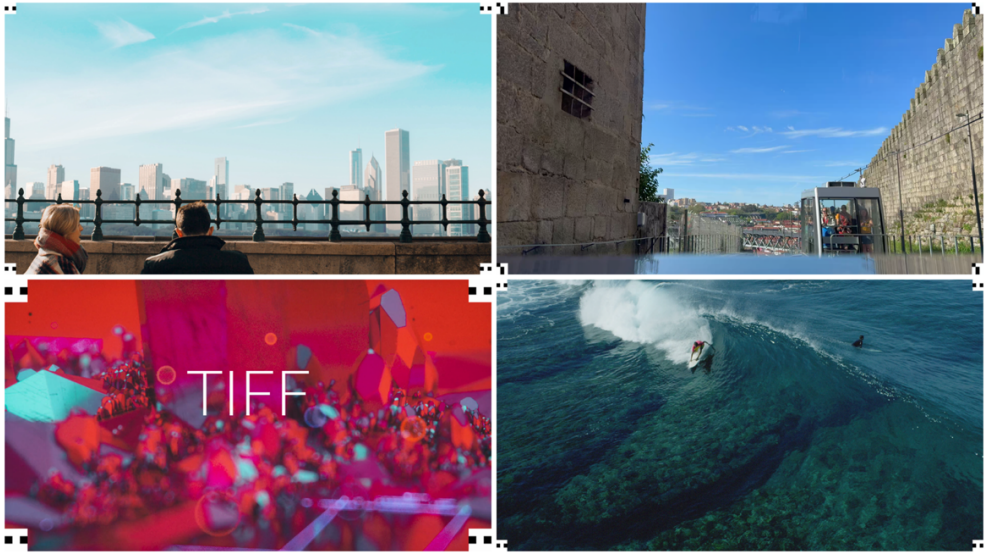

**Figure 2:** Examples of one frame from each of the four sequences in our dataset. Clockwise from top left: 2P5D, MovingTrainPorto, Waves, TiffText. Note the QR codes burned into the corner of the frames for verifying genlock in analysis of differences between reference footage and compressed versions.

## Our Dataset

We use 4 test video sequences shown in **Figure 2**. For each of the clips, we have burned a QR code into the 4 corners of the image to enable accurate sync between the recorded images and the reference on disk. The sequences are described in the table below.

| Name | Format | Duration (secs) | Description |
| --- | --- | --- | --- |
| 2P5D | UHD-1@8-bit (3840x2160) | 5 | Two people sitting with a train moving into shot. |
| MovingTrainPorto | UHD-1@10-bit (3840x2160) | 7 | Train with reflections and natural scenery. |
| TiffText | Full-HD@8-bit | 5 | Synthetic animated text with saturated colours. |
| Waves | DCI-4K@10-bit (4096x2160) | 7 | Complex scene of person surfing. |

# The Evaluation Process

The key target in this work is to evaluate the picture quality of some test picture formats rendered in-camera relative to the in-camera picture quality when the raw uncompressed data is rendered. Consider some example footage A for example. We first playback (at 30 FPS) the raw images (ARef) as TIF through the LED wall, and then record that in the camera as reference footage ARefCam. We then create a compressed version of A using some parameter settings (ADeg), and decode (ADegDec) and play that back on the LED wall, then capture in camera (denoted as ADegDecCam). The camera and screen setup are not changed between these two recordings so the environment is kept constant. The only change that would affect picture quality is therefore the difference between ARef and ADegDec.

To measure the impact of the compression format on in-camera quality we therefore measure the difference between ARefCam (the best possible quality possible in this system) and ADegDecCam. Since the camera viewpoint and screen position are unchanged between clip rendering, the differences are due purely to the display and camera recording. There is also no need to compensate for any motion or geometric distortion between captured ARefCam and ARefDegCam.

A flowchart of the process is shown in **Figure 3**. To provide monitoring of genlock, we insert a binary QR code (90x90 pixels) at the four corners of each rendered frame. This allows identification of the displayed frame in the camera record and hence detection of loss of genlock. Additionally, the input video is converted to a GPU encoder-friendly format with YUV 4:2:0 chroma subsampling. Video capture via a capture card outputs 12-bit UHD-1 ProRes at 4:4:4. That output ProRes file is used for calculating metrics. The encoder configurations that were tested are discussed next.

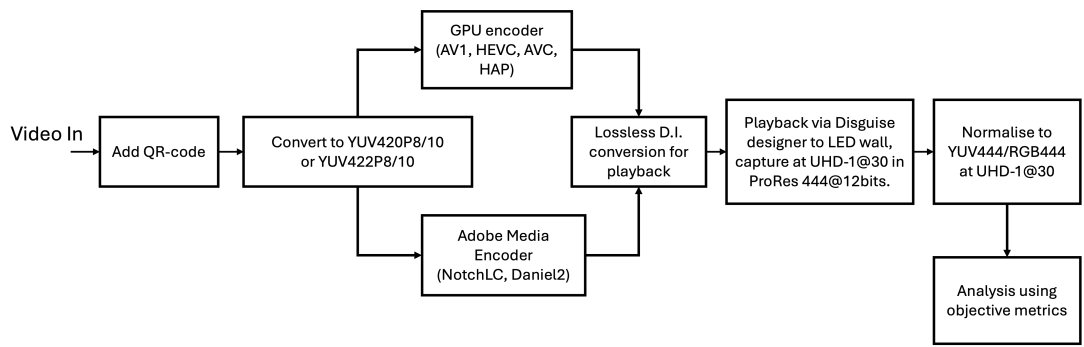

**Figure 3:** Flowchart summarizing our evaluation process. The Lossless DI stage exists only to facilitate playback in an existing VP setup because infrastructure for using arbitrary hybrid compressed bitstreams is not yet available.

## *Encoder Configuration*

All videos were encoded using a range of Quality Points controlled by Quantiser Parameter (QP) across two different GOP sizes, where applicable (i.e. not for Daniel2, NotchLC, HAP). The GOP sizes tested were All-Intra (GOP-0) and One I-Frame at the start (GOP -1). Additional GOP sizes of 1, 2 sec, and 5 frames were tested but yielded similar results w.r.t. quality.

For each codec, five QP points were selected, representing approximately one Just-Noticeable-Difference (JND) in VMAF for each video, with a minimum VMAF score of 82. We note that i) NVIDIA's HEVC codec could not exceed 500 Mb/s in Constant Bitrate Mode (CBR), ii) Daniel2 could not achieve bitrates as low as other codecs, iii) NVIDIA's H.264 did not support 10-bit encoding, and iv) NotchLC and HAP lacked rate control and target bitrate options. Our final parameterisations for each codec took these observations into account and are shown in the table below.

| Codec | Implementation | Rate-Control Mode |
|---|---|---|
| H.264/AVC | NVIDIA Encoder (40 Series) | $QP \in \{11..51\}$ |
| H.265/HEVC | NVIDIA Encoder (40 Series) | $QP \in \{11..50\}$ |
| AV1 | NVIDIA Encoder (40 Series) | $QP \in \{15..230\}$ |
| HAP | FFmpeg Reference Implementation | HAP-Q mode |
| NotchLC | Adobe Media Convert | Constant Quality 4 Quality Levels $\in$ {good, excellent, optimal, best} |
| Daniel2 | Adobe Media Convert | Constant Quality CQ values of $\in \{20..95\}$ |

*Environment configuration*

The experiment was conducted at Disguise Studio in London, UK, using an LED wall (4.5m x 2m) and a Red Komodo 6K camera with a Fujinon MK18-55 lens. The video infrastructure captures UHD-1 (3840x2160) at 30 FPS using AJA Ki Pro Ultra 12G. The system is genlocked to 30 FPS. Three VP environments were tested:

i) Set Extension: The full LED screen was used, but a portion of the video was projected in the real world. An example is shown in **Figure 4**, where the embedded real-world portion has a different colour and tone, indicated by the mask on the right.
ii) Full Screen: The entire video was displayed on the full LED wall.
iii) Zoom-in: A portion of the LED wall was used as illustrated in **Figure 5**.

In practice we find that set-extension and Zoom-in modes are more or less identical because we only analyse that portion of the set-extension imaged on the LED wall. Also, as we in fact render our zooms on almost the full wall, ii) and iii) are also identical. So we only report on zoom modes in this paper.

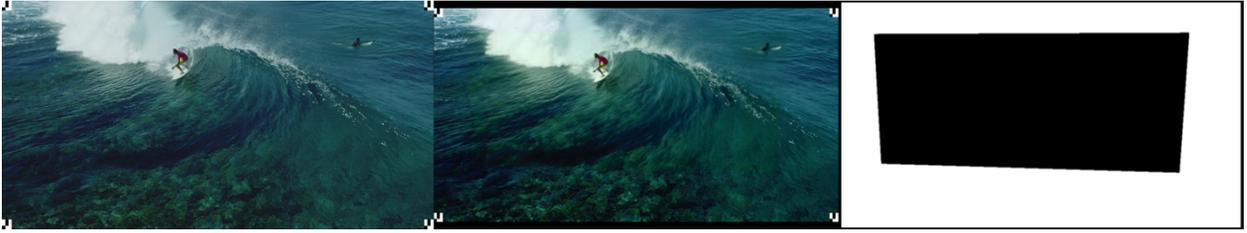

**Figure 4:** Set Extension mode used for evaluation. The Original frame (left) is rendered in Set-Extension mode (middle), with the alpha mask of the Set-Extension shown on the right.

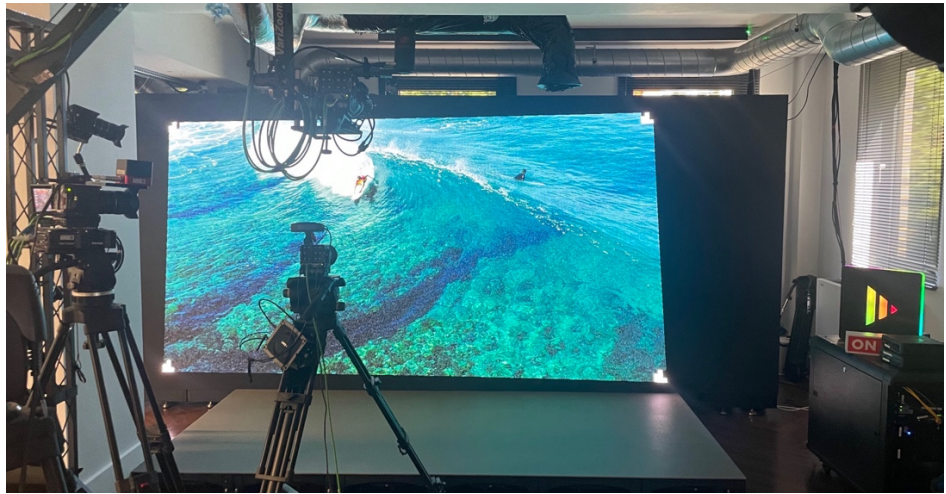

**Figure 5:** Zoomed-in VP test case is shown here in which a part of the screen is used for the capture. In this mode aliasing artefacts on the rendition in-camera are probably most likely.

## Results and Discussion

We first evaluate the theoretical best performance of these encoders outside of the VP setup. Then we go on to evaluate performance in the various VP modes. Note that we only compare H.264 on the 2P5D and TiffText sequences as they are 8 bit while the other sequences are 10 bit.

**Theoretical Limit of Performance. Figure 6** shows the performance of our codecs tested using the 2P5D and Waves sequences. In this case we are comparing the raw performance of the encoders on disk (ADegDec) before rendering through our VP setup. This is the theoretical limit of performance that we would expect if the VP setup itself caused no degradation. To allow a fair comparison with NotchLC, Daniel2 and HAP we turn off the interframe prediction in the hybrid codecs and show here only GOP length 0 encoding i.e. intra-frame only. Considering PSNR relative to bitrate and assuming that 50dB PSNR is virtually lossless, we see that all of our hybrid codecs can achieve 50dB or better while NotchLC, Daniel 2 cannot. For the Waves sequence which is more textured NotchLC can indeed achieve 50dB. VMAF metrics show a different picture. VMAF is better correlated with human perception, and here we see that in both sequences all the codecs can perform well, achieving ~100 VMAF for certain bitrate ranges. Significantly however, AV1 and HEVC can achieve ~100 VMAF at less than 200 Mbps for the 2P5D sequence while NotchLC, HAP operate in the ~800 Mbps range to achieve that score. Daniel 2 performs much better but still at least 300 Mbps range. Colour performance measured using CVVDP shows that all the codecs achieve a very high level of colour integrity, scoring between 9.5 and 10 for all sequences but again the hybrid codecs attain this high level of performance at a significantly lower bitrate than NotchLC or HAP.

**Zoom Mode.** Moving on to performance through the VP setup (quality of ADegDecCam measured w.r.t. ARefCam), our evaluation results on this same pair of clips are shown in **Figure 7** using GOP-0 (Intra-only) mode for the hybrid codecs. Also included in **Figure 8** is the comparison using a single GOP (GOP -1) for the whole clip. In all cases, the separation between the performance of the intermediate codecs NotchLC, HAP and Daniel2 is more stark.

Compared to NotchLC and HAP, the hybrid codecs can achieve the same quality at an order of magnitude less bitrate (x10) depending on the chosen QP setting and content. For the simpler content of 2P5D, x10 is possible, while for Waves (with more texture) the benefit is in the x2-x5 range. All the codecs appear to achieve better quality than Daniel2.

We were surprised in these experiments to find that the performance of the VP setup through the camera with H.264 is in the ballpark (on a log scale of bitrate) of the performance of AV1 and HEVC. It is also interesting that the upper ranges of VMAF and PSNR are about 10-20% less than the theoretical maximum possible in **Figure 6** regardless of the encoder. This is because our reference footage is now ARefCam which contains degradations due to playback and the nature of the VP setup. These degradations are random and so even if we compared ARefCam with another recording of ARefCam at a different time, we would not see a perfect reproduction. We measured this with the TiffText sequence by comparing ARefCam played back 50 times consecutively. The maximum PSNR observed was not 50dB but 37 dB within a very small range of 0.02 dB. Hence the maxima in Figures 7, 8 and 9 are indeed the maximum measurable quality as expected. These points show that in fact, the bigger limiting factor in picture quality for VP is the LED Panel rendering and the aliasing between the panel and the camera capture, not so much the hybrid codec choice.

To summarise the impact of our hybrid codecs on quality/bitrate we observe that NotchLC seems to generate the highest objective quality at the highest bitrate across all encoders. Using NotchLC as the quality reference we measure the minimum bitrate generated by each hybrid encoder in order to attain VMAF score of 90. The tables below report on the bitrate savings (in multiples of the NotchLC bitrate) compared to the NotchLC reference for two different GOP settings.

| Bitrate Savings using NotchLC as reference (Bitrate Mb/s) using All-Intra | | | | | |
|---|---|---|---|---|---|
| Codec/Video | MovingTrainPorto (864.1) | Waves (1449.1) | Tifftext (281.9) | 2P5D (772.8) | Average (842.0)r |
| HEVC | 3.64x (237.5) | 1.28x (1135.2) | 13.44x (21.0) | 31.31x (24.7) | 12.42x |
| AV1 | 4.02x (215.2) | 1.23x (1181.5) | 10.61x (26.6) | 28.11x (27.5) | 10.99x |
| H.264 | N/A | N/A | 8.59x (32.8) | 13.75x (56.2) | 11.17x |
| HAP | 0.36x (2394.9) | 0.73x (1987.9) | 0.66x (425.7) | 0.83x (932.7) | 0.65x |
| Daniel2 | N/A | 0.99x (1467.5) | N/A | 3.57x (216.7) | 2.28x |

| Bitrate Savings using NotchLC as reference (Bitrate Mb/s) using GOP -1 | | | | | |
|---|---|---|---|---|---|
| Codec/Video | MovingTrainPorto (864.1) | Waves (1449.1) | Tifftext (281.9) | 2P5D (772.8) | Average (842.0) |
| HEVC | 12.37x (69.9) | 1.31x (1109.9) | 19.25x (14.6) | 195.66x (4.0) | 57.2x |
| AV1 | 13.71x (63.1) | 1.34x (1084.5) | 23.14x (12.2) | 233.06x (3.3) | 67.8x |
| H.264 | N/A | N/A | 18.86x (15.0) | 129.77x (6.0) | 74.3x |

The tables show an average gain of about 12x in bitrate savings for HEVC or AV1 in Intra only mode. Daniel2 and H.264 cannot attain VMAF 90 across all sequences. Where this is achieved Daniel2 and HAP are often only able to do this at higher bitrate than NotchLC especially for complex content e.g. Waves.

**Figure 8** shows bitrate savings using hybrid codecs with just one intra frame in the sequence i.e. using the motion prediction mode of the encoders. Dramatically improved savings are now possible for no loss of objective quality e.g. for 2P5D we approach 230x savings (see tables) but more complex content (Waves) is less affected. **Figure 9** shows the same evidence for MovingTrainPorto and TiffText sequences.

**Figure 10**: gives us good confidence that all of the encoders investigated here preserve colour well through the VP setup as CVVDP is at least 9.4 out of 10 max.

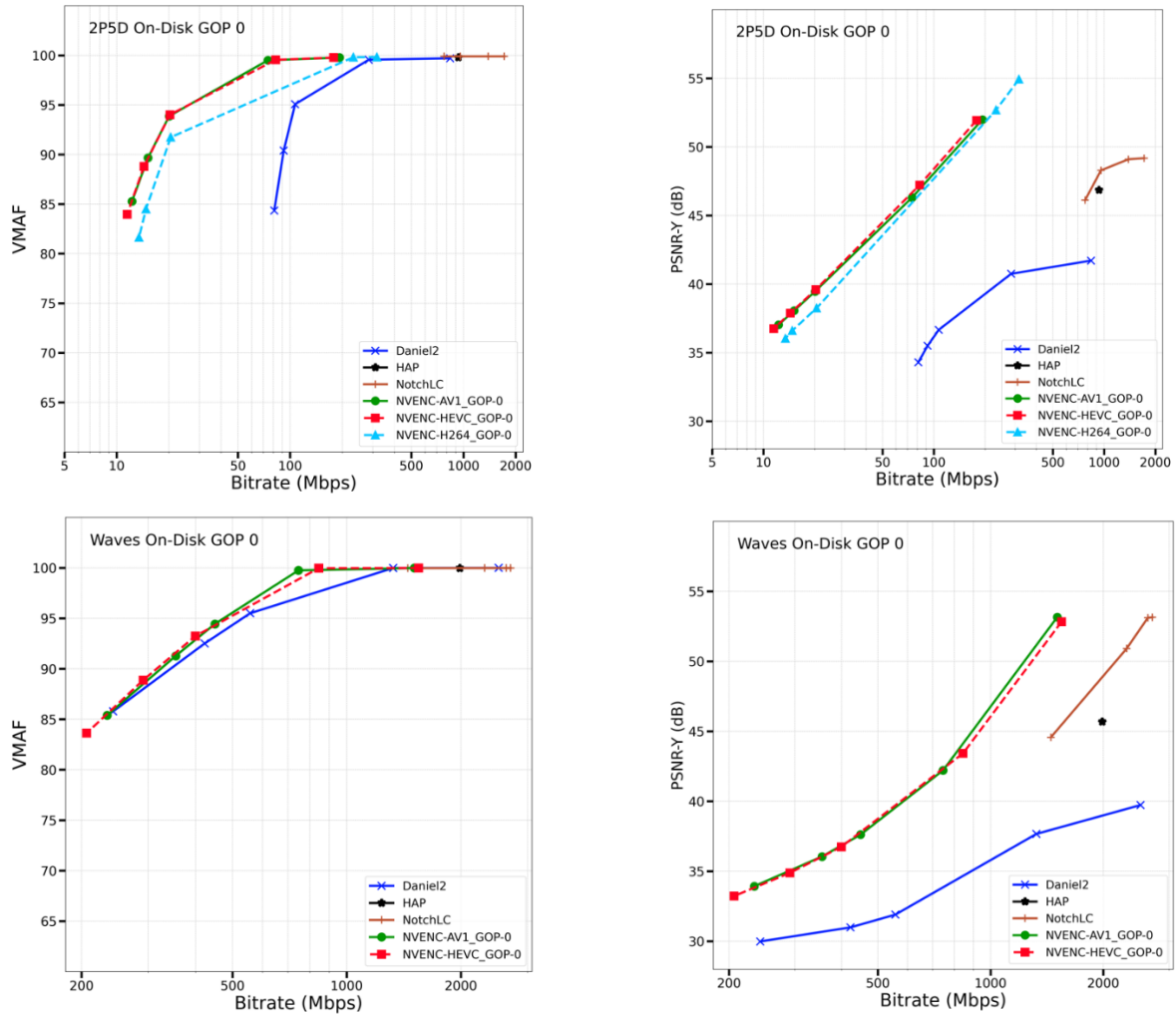

**Figure 6:** Rate/Quality curves measuring VMAF (left) and PSNR (right) as a function of bitrate for all of our codecs tested across various parametrisations. Top row: 2P5D, Bottom row Waves Test sequence. In this case all our hybrid codecs H264/HEVC/AV1 used Intra only encoding. The best performing systems appear to the top left of these plots. As can be seen the hybrid codecs perform better in PSNR or the same in VMAF at much lower bitrates than HAP, Notchlc or Daniel2.

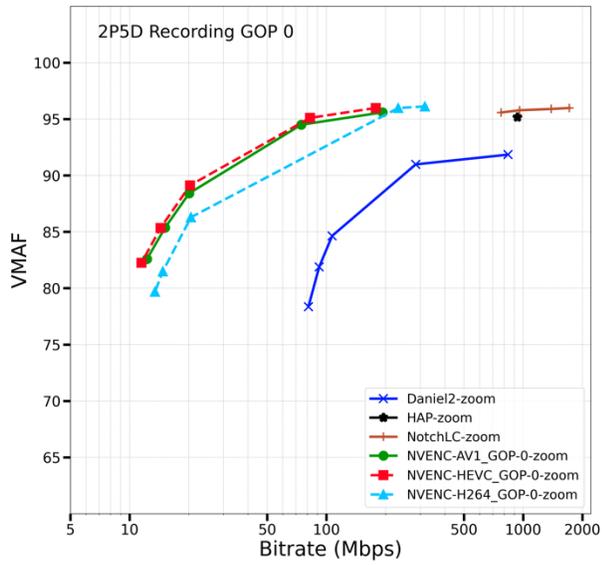
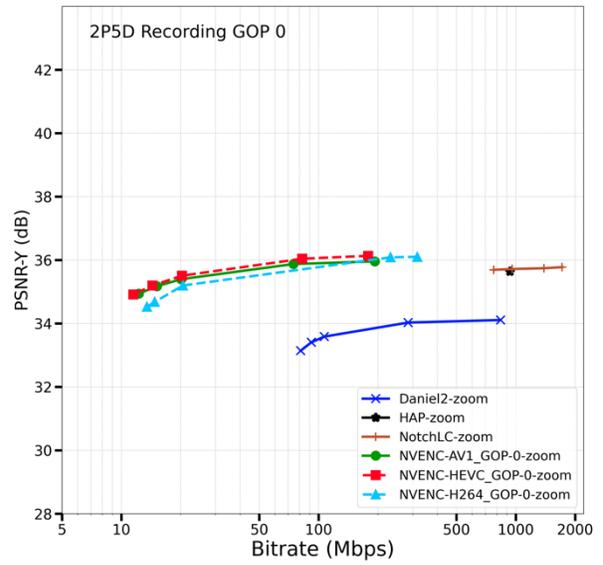
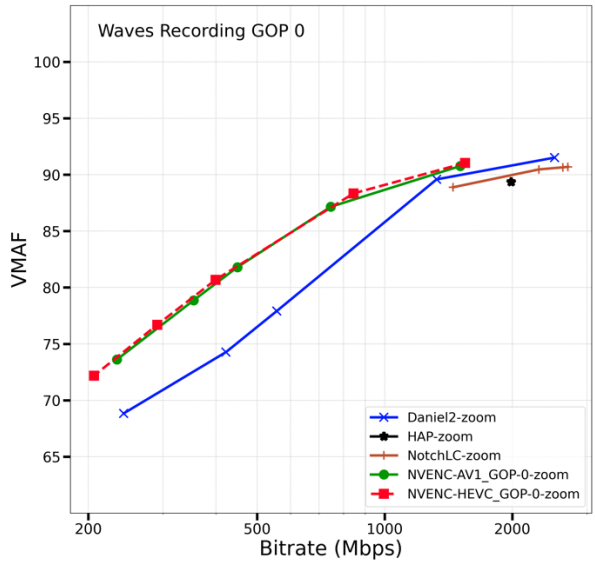
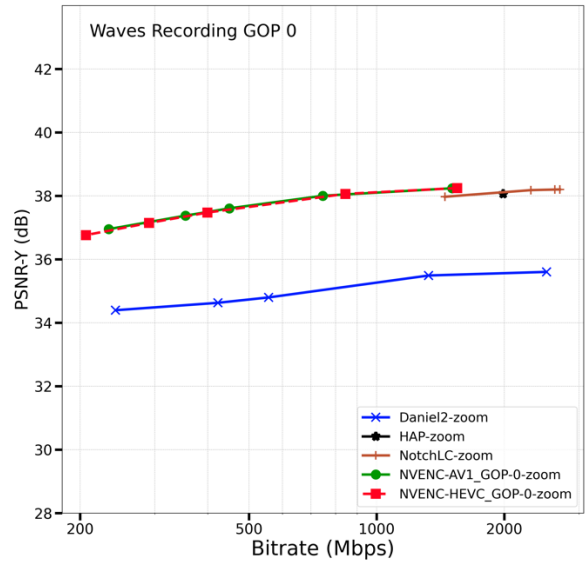

**Figure 7:** Evaluation of performance of VP setup in zoom mode using 2P5D (top) and Waves (bottom) sequences, with VMAF (left) and PSNR (right) columns. The Hybrid codecs used GOP-0 (intra only) modes here. Hybrid codecs achieve high quality at much lower bitrates.

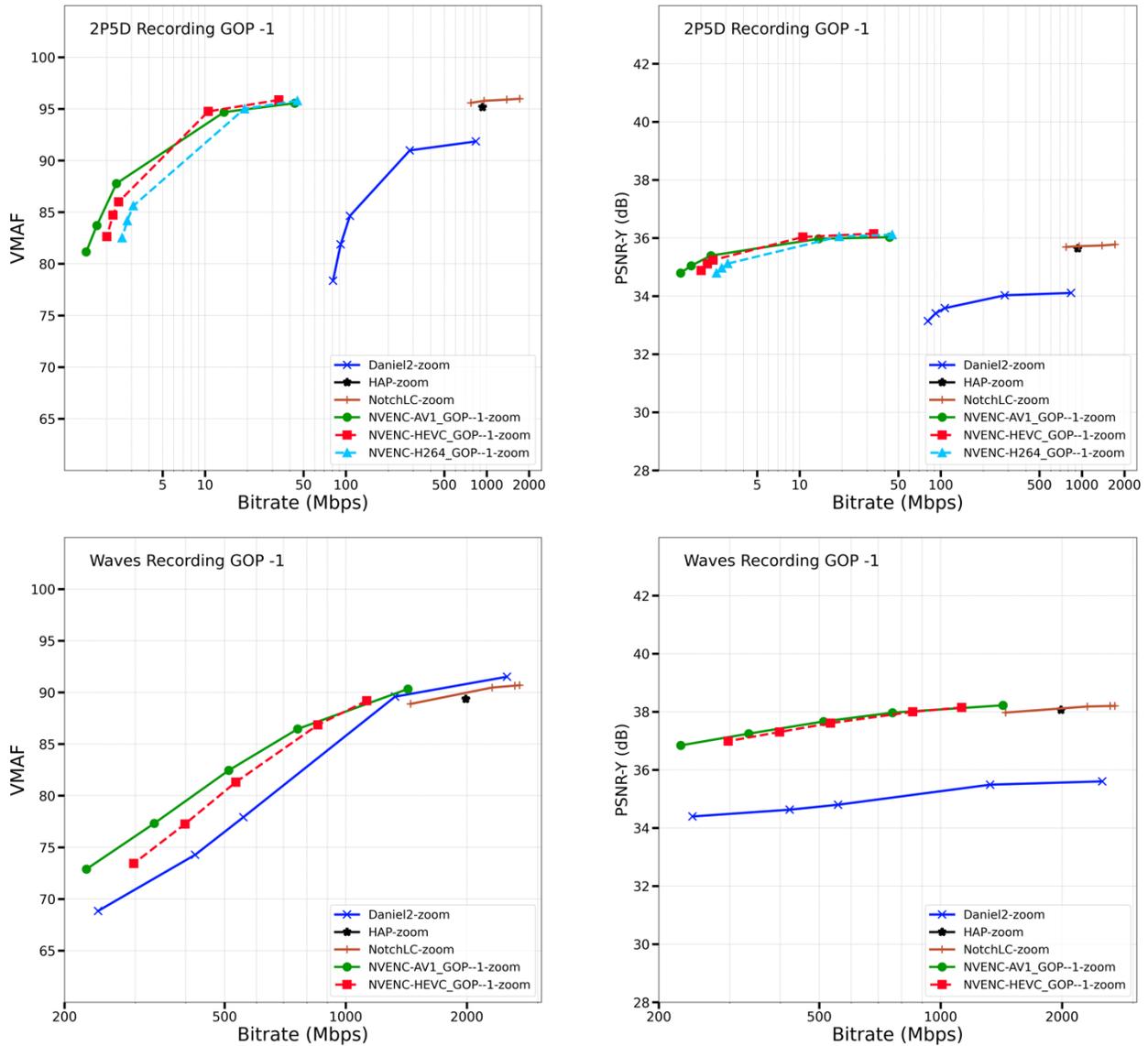

**Figure 8:** Evaluation of performance of VP setup in zoom mode using 2P5D (top) and Waves (bottom) sequences, with VMAF (left) and PSNR (right). These all show performance with one Intra frame per sequence (GOP -1). This yields strong reduction in bitrate in the hybrid codecs with negligeable impact on quality.

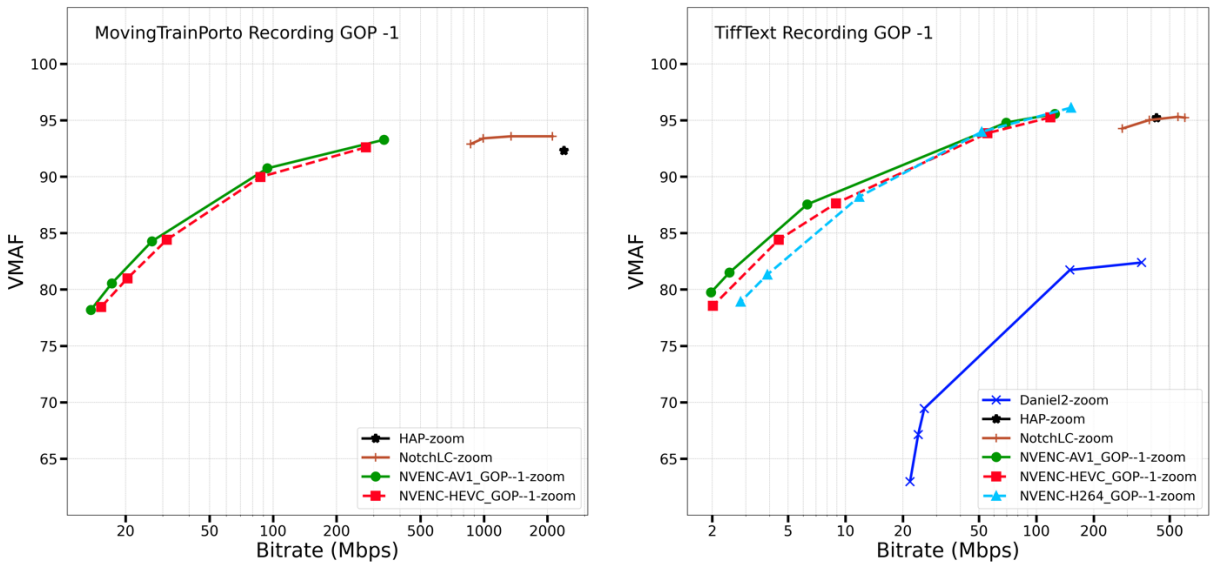

**Figure 9**: Evaluation as in Figure 8 (VMAF only) for MovingTrainPorto and TiffText sequences with one Intra-frame GOP.

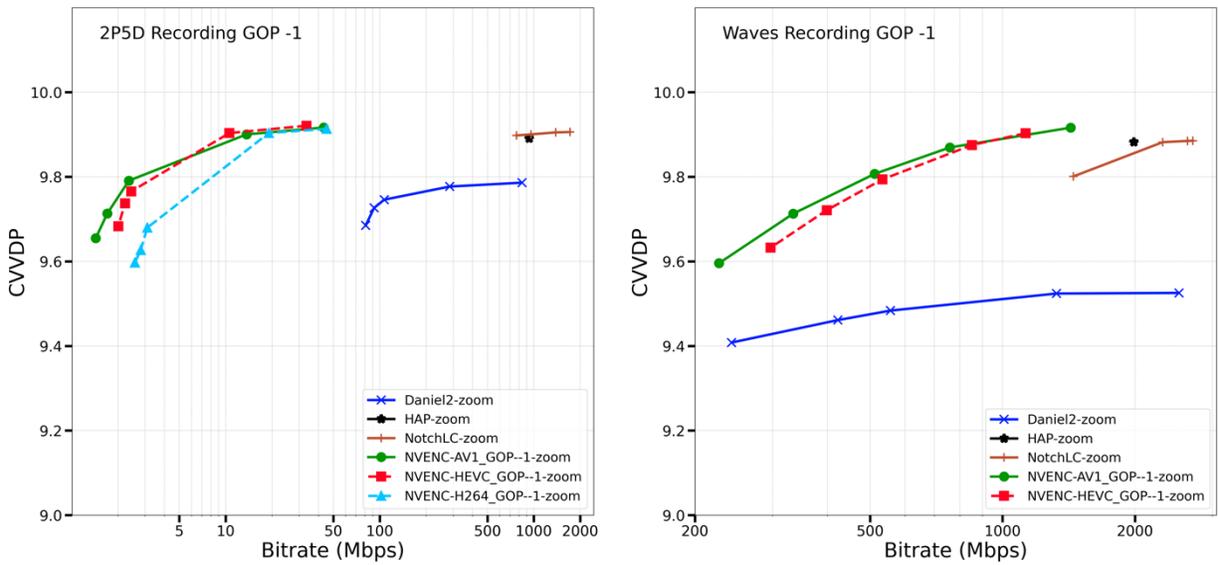

**Figure 10:** CVVDP for one Intra frame in the sequence measured for 2P5D (left) and Waves (right) in Zoom mode. Colour preservation is very good across all codecs but again hybrid codecs perform best.

## Conclusion

In this paper we have presented a methodology for analysis of picture quality in VP. Using our particular VP setup, we show that using hybrid codecs can achieve the same quality in-camera at orders of magnitude less bitrate than intermediate codecs. Surprisingly, when using the NVIDIA's "Nvenc" toolbox which implements several encoders in hardware, we can achieve better encoding FPS than for the standard intermediate codecs used in VP. These results show that it is definitely worth deploying an infrastructure that can employ hybrid codecs within VP setups. The fact that hardware encoders already exist alongside GPU devices makes for feasible deployment.

Based on our measurement with perceptual metrics we can cautiously recommend both HEVC and AV1 as viable alternatives to existing streams to LED panels. Using Intra-only modes is safe from the point of view of latency but if some latency can be tolerated then 5 second GOPs will bring even more gains. Our next steps involve a consideration of subjective testing in this environment. The challenge there is capturing the very subtle differences between the encoded representations which we have noticed to be barely visible at these high bitrates.

## Acknowledgements

This work was funded by the Horizon CL4 2022 - EU Project Emerald – 101119800, ADAPT-SFI Research Center, Ireland with Grant ID 13/RC/2106_P2, and YouTube & Google Faculty Awards.

## Bibliography

[1] G. W. Perkins and S. Echeverry, "Virtual Production in Action: A Creative Implementation of Expanded Cinematography and Narratives," in *ACM SIGGRAPH Posters*, 2022.

[2] T. Zhang, "From digital visual effects to emerging in-camera visual effects: investigating the change of workflow, occupational roles and common challenges in Southeast Asian and East Asian countries.," *Nanyang Technological University ,* 2024.

[3] Disguise, Welcome to the New Era of Film: Five ways virtual production is revolutionising filmmaking, Disguise, 2023.

[4] N. Kadner, The Virtual Production Field Guide Volume 1, Epic Games, 2019.

[5] P. Debevec and C. LeGendre, "HDR Lighting Dilation for Dynamic Range Reduction on Virtual Production Stages.," in *ACM SIGGRAPH 2022 Posters*, 2022.

[6] R. Hendricks, "Rapid Industry Solutions (RiS) Initiative, On-Set Virtual Production (OSVP), and Open Services Alliance (OSA)," *SMPTE Motion Imaging Journal,* vol. 8, no. 132, pp. 78-80, 2023.

[7] B. Desowitz, "Reverse Engineering 'Gravity'," Animation World Network, 4 October 2013. [Online]. Available: https://www.awn.com/vfxworld/reverse-engineering-gravity. [Accessed 03 August 2024].

[8] S. Dalkian, S. Miglio and others, "nDisplay Technology: Limitless scaling of real-time content," *Technical Whitepaper,* pp. 1-24, 2019.

[9] Disguise, "Eight benefits of virtual production," Disguise, April 2021. [Online]. Available: https://www.disguise.one/en/insights/blog/eight-benefits-of-virtual-production/. [Accessed 05 August 2024].

[10] M. Kavakli and C. Cremona, "The virtual production studio concept–an emerging game changer in filmmaking," *IEEE Conference on Virtual Reality and 3D User Interfaces (VR),* pp. 29-37, 2022.

[11] F. Pires, R. Silva and R. Raposo, "A survey on virtual production and the future of compositing technologies," *Avanca Cinema Journal,* vol. 21, no. 692-9, 2022.

[12] Tom Whittock, "The Key to Superior Content Rendering," Disguise, 12 2023. [Online]. Available: https://www.disguise.one/sites/default/files/2023-12/Renderstream_Whitepaper.pdf. [Accessed 05 August 2024].

[13] M. Henrique and G. Sullivan, "YCoCg-R: A color space with RGB reversibility and low dynamic range," ISO/IEC JTC1/SC29/WG11 and ITU-T SG16 Q6, 2003.

[14] Cinegy , "Cinegy Daniel2 GPU Codec Enters Public Beta," Cinegy, 6 June 2017. [Online]. Available: https://www.cinegy.com/news/cinegy-daniel2-gpu-codec-enters-public-beta/. [Accessed 06 August 2024].

[15] Kostya, "A cursory glance at Daniel codecs," 10 July 2022. [Online]. Available: https://codecs.multimedia.cx/2022/07/a-cursory-glance-at-daniel-codecs/. [Accessed 06 August 2024].

[16] N. Porrmann, "10bit Workflow Viability + Codec Comparison," Dandelion-burdock, June 2019. [Online]. Available: https://dandelion-burdock.com/articles/10bit-workflow-viability. [Accessed 6 August 2024].

[17] K. Seshadrinathan, R. Soundararajan, A. C. Bovik and L. K. Cormack, "Study of subjective and objective quality assessment of video," *IEEE transactions on Image Processing,* vol. 19, no. 6, pp. 1427-1441, 2010.

[18] V. Vibhoothi, A. Katsenou, F. Pitie, K. Domijan and A. Kokaram, "Subjective Assessment of the Impact of a Content Adaptive Optimiser for Compressing 4K HDR Content With AV1," in *2023 IEEE International Conference on Image Processing (ICIP),*, Kuala Lumpur, Malaysia,, 2023.

[19] K. Murray and V. Simion, Color Management in Virtual Production, vol. 20, Los Angeles: ACM SIGGRAPH, 2023.

[20] J. Y. Lin, E. C.-H. Wu, C.-C. J. Kuo, P. L. Callet, T. Goodall and A. Aaron, Toward A Practical Perceptual Video Quality Metric, Los Gatos: Netflix Technical Blog, 2016.

[21] Z. Li, C. Bampis, J. Novak, A. Aaron, K. Swanson, A. Moorthy and J. De Cock, "VMAF: The Journey Continues," Netflix, 25 October 2018. [Online]. Available: https://netflixtechblog.com/vmaf-the-journey-continues-44b51ee9ed12. [Accessed 11 August 2024].

[22] R. K. Mantiuk, P. Hanji, M. Ashraf, Y. Asano and A. Chapiro, "ColorVideoVDP: A visual difference predictor for image, video and display distortions," *ACM Trans. Graph,* vol. 43, no. 4, p. 20, 2024.